\documentclass[11pt,letterpaper]{llncs}
\usepackage{amsmath}
\usepackage{amssymb}
\usepackage{graphicx}
\usepackage{marvosym}
\usepackage{url}
\usepackage{fancyhdr}

\usepackage{geometry}
\newcommand{\keywords}[1]{\par\addvspace\baselineskip
\noindent\keywordname\enspace\ignorespaces#1}

\fancypagestyle{firstpage}{\fancyhf{}
\setcounter{page}{1}
\fancyhead[C]{\small{International Journal of Computer Graphics $ \& $ Animation (IJCGA) Vol.9, No.1/2/3, July 2019
}}
\fancyfoot[L]{DOI: 10.5121/ijcga.2019.9301}
\rfoot{\thepage}
}

\usepackage{listings}

\begin{document}

%\mainmatter  % start of an individual contribution

\title{\LARGE{Another Simple but Faster Method for 2D Line Clipping}}

%\titlerunning{Lecture Notes in Computer Science: Authors' Instructions}

\author{\large{Dimitrios Matthes \and Vasileios Drakopoulos}}
\institute{\large{Department of Computer Science and Biomedical Informatics, \\School of Science, University of Thessaly,\\ Lamia 35131, Greece}}

%\author{Alfred Hofmann%
%\thanks{Please note that the LNCS Editorial assumes that all authors have used
%the western naming convention, with given names preceding surnames. This determines
%the structure of the names in the running heads and the author index.}%
%\and Ursula Barth\and Ingrid Haas\and Frank Holzwarth\and\\
%Anna Kramer\and Leonie Kunz\and Christine Rei\ss\and\\
%Nicole Sator\and Erika Siebert-Cole\and Peter Stra\ss er}

%\institute{Springer-Verlag, Computer Science Editorial,\\
%Tiergartenstr. 17, 69121 Heidelberg, Germany\\
%\url{http://www.springer.com/lncs}}


%\toctitle{Lecture Notes in Computer Science}
%\tocauthor{Authors' Instructions}

\maketitle

\thispagestyle{firstpage}

\begin{abstract}

The majority of methods for line clipping make a rather large number of comparisons and involve a lot of calculations compared to modern ones. 
Most of the times, they are not so efficient as well as not so simple and applicable to the majority of cases.
Besides the most popular ones, namely, Cohen-Sutherland, Liang-Barsky, Cyrus-Beck and Nicholl-Lee-Nicholl, other line-clipping methods have been presented over the years, each one having its own advantages and disadvantages. 
In this paper a new computation method for 2D line clipping against a rectangular window is introduced. 
The proposed method has been compared with the afore-mentioned ones as well as with two others; namely, Skala and Kodituwakku-Wijeweera-Chamikara, with respect to the number of operations performed and the computation time.
The performance of the proposed method has been found to be better than all of the above-mentioned methods and it is found to be very fast, simple and can be implemented easily in any programming language or integrated development environment.
\keywords{2D Computer Graphics; Computer Graphics Education; Geometry; Line Clipping; Programming Education; Computation method}
\end{abstract}  

%\begin{abstract}
%The abstract should summarize the contents of the paper and should
%contain at least 70 and at most 150 words. It should be written using the
%\emph{abstract} environment.
%\keywords{We would like to encourage you to list your keywords within
%the abstract section}
%\end{abstract}

\section{Introduction and historical background}

In computer graphics, any procedure that eliminates those portions of a picture that are either inside or outside a specified region of space is referred to as a {\em clipping algorithm} or simply {\em clipping}. 
The region against which an object is to be clipped is called a {\em clipping object}.
In two-dimensional clipping, if the clipping object is an axis-aligned rectangular parallelogram, it is often called the {\em clipping window} or {\em clip window}.  
Usually a clipping window is a rectangle in standard position, although we could use any shape for a clipping application.
For a three-dimensional scene it is called a {\em clipping region}; see \cite{HBC14}.
\begin{figure}[htb]
\centering
\includegraphics[width=.8\linewidth]{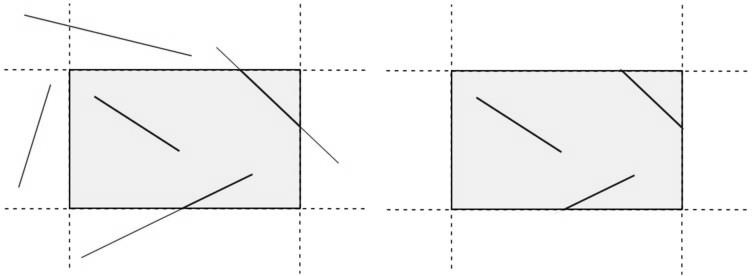}
\parbox[t]{.9\columnwidth}{\relax}
\caption{\label{fig:clippingBeforeAfter}
Region before (left) and after (right) 2D line clipping.}
\end{figure}

{\em Line clipping} is the process of removing lines or portions of lines outside an area of interest. 
Typically, any line or part thereof which is outside of the viewing area is removed (Fig.~\ref{fig:clippingBeforeAfter}). 
Most of the times, this process uses mathematical equations or formulas for removing the unecessary parts of the line. 
The programmer draws only the part of the line which is visible and inside the desired region by using, for example, the slope-intercept form $y=m x + b$, where $m$ is the slope or gradient of the line, $b$ is the $y$-intercept of the line and $x$ is the independent variable of the function $y = f(x)$ or just the vector equation.
The most common application of clipping is in the viewing pipeline, where clipping is applied to extract a designated portion of a scene (either
two-dimensional or three-dimensional) for display on an output device. 
Clipping methods are also used to antialias object boundaries, to construct objects using solid-modelling methods, to manage a multiwindow environment and to allow parts of a picture to be moved, copied, or erased in drawing and painting programs; see for example \cite{HB94} or \cite{Dim15}. 

There are four primary methods for line clipping: Cohen-Sutherland, Cyrus-Beck \cite{CB78}, Liang-Barsky \cite{LB84} and Nicholl-Lee-Nicholl \cite{NLN87}. 
Over the years, other algorithms for line clipping appeared, like Fast Clipping \cite{SPY87}, D\"{o}rr \cite{Dor90}, Andreev and Sofianska \cite{AS91}, Day \cite{Day92}, Sharma and Manohar \cite{SM92}, Skala \cite{Ska93} \cite{Ska94} \cite{Ska05}, Slater and Barsky \cite{SB94}, Ray \cite{Ray12}, \cite{Ray12a} but many of them are variations of the first two ones; see \cite{PJ13} or \cite{KPK18} for a different approach. 
In general, the existing line-clipping algorithms can be classified into three types: the encoding approach (with the Cohen-Sutherland algorithm as a representative), the parametric approach (with the Liang-Barsky and the Cyrus-Beck algorithms as representatives) and the Midpoint Subdivision algorithms.

\begin{figure}[htb]
\centering
\includegraphics[width=.8\linewidth]{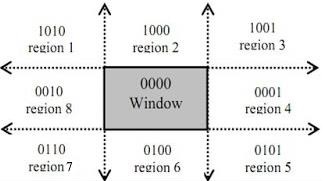}
\parbox[t]{.9\columnwidth}{\relax}
\caption{\label{fig:cohenSutherland}The nine regions of the Cohen-Sutherland algorithm in the 2D space.}
\end{figure}

The algorithm of Danny Cohen and Ivan Sutherland was developed in 1967 during the making of a flight simulator. 
It is considered to be one of the first line-clipping algorithms in the computer graphics history. 
According to this, the 2D space in which the line resides is divided into nine regions. 
The algorithm determines first in which regions the two points that define the line are in and then performs complete, partial or no drawing of the line at all; see for example \cite{FDFH90}, p. 113 or \cite{GG08}  (Fig.~\ref{fig:cohenSutherland}).  
The method that is used to decide if a line is suitable for clipping or not performs logical AND operation with the {\em region codes} or {\em outcodes} of the line end points. 
After the logical AND, if the result is not 0000, the line is completely outside the clipping region \cite{IMM11}. 
This technique is also referred to as {\em Encoding and Code Checking} in \cite{LWP02}.

The method of Mike Cyrus and Jay Beck is a general line-clipping algorithm, but it introduces extra floating point operations for determining the value of a parameter corresponding to the intersection of the line to be clipped with each window edge \cite{KEC90}. 
It is of $O(N)$ complexity, where $N$ is a number of facets, and is primarily intended for clipping a line in the parametric form against a convex polygon in two dimensions or against a convex polyhedron in three dimensions.

Midpoint subdivision algorithm is an extension of the Cyrus-Beck algorithm and follows a divide and conquer strategy. 
It is mainly used to compute visible areas of lines that are present in the view port of the sector or the image. 
It follows the principle of the bisection method and works similarly to the Cyrus-Beck algorithm by bisecting the line into equal halves. 
But unlike the Cyrus-Beck algorithm, which only bisects the line once, Midpoint Subdivision Algorithm bisects the line numerous times.
The Midpoint Subdivision algorithm is not efficient unless it is implemented in hardware. 

\begin{figure}[htb]
\centering
\includegraphics[width=.8\linewidth]{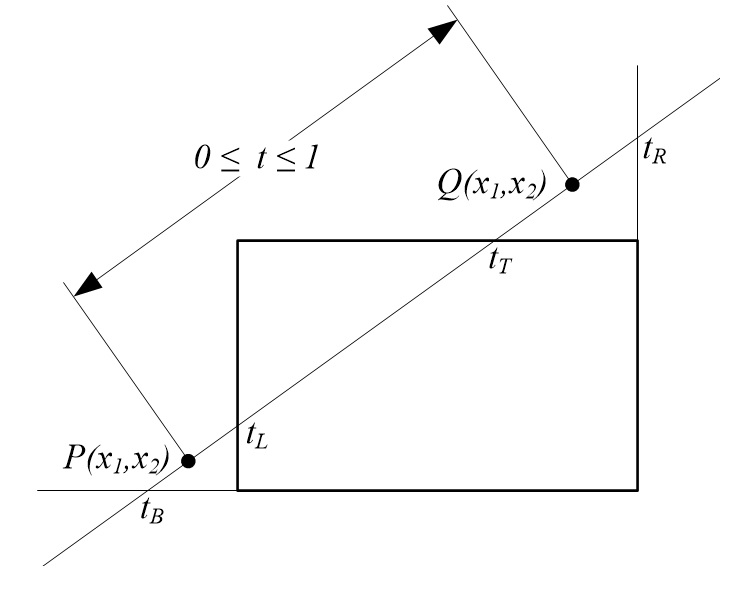}
\parbox[t]{.9\columnwidth}{\relax}
\caption{\label{fig:liangBarsky}Defining the line for clipping with the Liang-Barsky algorithm.}
\end{figure}

On the other hand, You-Dong Liang and Brian Barsky have created an algorithm that uses floating-point arithmetic for finding the appropriate end points with at most four computations \cite{Nis17a}. 
This algorithm uses the parametric equation of the line and solves four inequalities to find the range of the parameter for which the line is in the viewport \cite{LB84}.
The method of  Liang-Barsky is very similar to Cyrus-Beck line-clipping algorithm. 
The difference is that Liang-Barsky is a simplified Cyrus-Beck variation that was optimised for a rectangular clip window; see Fig.~\ref{fig:liangBarsky}.
In general, the Liang-Barsky algorithm is more efficient than the Cohen-Sutherland line-clipping algorithm.

The Nicholl-Lee-Nicholl algorithm is a fast line-clipping algorithm that reduces the chances of clipping a single line segment multiple times, as may happen in the Cohen-Sutherland algorithm. 
The clipping window is divided into a number of different areas, depending on the position of the initial point of the line to be clipped.

The algorithm of Skala~\cite{Ska05} is based on homogeneous coordinates and duality.
It can be used for line or line-segment clipping against a rectangular window as well as against a convex polygon. 
The algorithm is based on classifying a vertex of the clipping window against a half-space given by a line $p: ax + by + c = 0$. 
The result of the classification determines the edges intersected by the line $p$. 
The algorithm is simple, easy to implement and extensible to a convex window as well. 
The line or line segment $p$ can be computed from points $r_1, r_2$ given in homogeneous coordinates directly using the cross product as
\[
  p = r_1\times r_2 = (x_1, y_1, w_1) \times (x_2, y_2, w_2)
\] 
or as 
\[
  p = r_1\times r_2 = (x_1, y_1, 1) \times (x_2, y_2, 1).
\] 

In 2013, a fast line clipping algorithm with slightly different approach from the above ones was introduced by Kodituwakku-Wijeweere-Chamikara \cite{KWC13}. 
It is newer and performs better than the Cohen-Sutherland and Liang-Barsky algorithms. 
It checks every boundary of the clipping area (top, bottom, left, right) and performs line clipping by using the equation of the line. 
Moreover, it checks if the line segment is just a point or parallel to principle axes. 

Depending on the programming language or integrated development environment, the implementation of each algorithm varies in speed. 
For example, the implementation of the Cohen-Sutherland algorithm in Scratch requires a relatively large number of bitwise AND operations for determining the regions that the line resides. 
However, bitwise AND is not embedded in Scratch so the programmer should have to create a function for this task which greatly impedes the algorithm. 
Sometimes, it is difficult to implement or teach to others a line-clipping algorithm because it uses either advanced mathematical concepts, like the Liang-Barsky, or it is complicated by using many conditions and comparisons (if..then..else..) like the Kodituwakku-Wijeweere-Chamikara.

The difficulties of the previous line-clipping algorithms seem to be overcomed by the proposed one; see also \cite{MD19}. 
Although it uses the main concept of the Kodituwakku-Wijeweere-Chamikara algorithm, it avoids many unecessary comparisons like the parallel lines or the dots. 
It aims at simplicity and speed and does only the necessary calculations in order to determine whether the beginning as well as the end of the line are inside the clipping region. 
Moreover, it uses the minimum, for all the tested algorithms, number of variables.

This article has the following structure. 
Section~\ref{Sec: proposed} presents the proposed line-clipping method, Section~\ref{Sec:Eval} presents the results after comparing the proposed with six other line-clipping algorithms (Cohen-Sutherland, Liang-Barsky, Cyrus-Beck, Nicholl-Lee-Nicholl, Skala and Kodituwakku-Wijeweere-Chamikara) all implemented in C++ with OpenGL, Section~\ref{Sec:Conclusions} presents the conclusions derived from the study and usage of the algorithm in practice as well as suggestions for improvement and, finally, Section~\ref{Sec:Sum} summarises all the findings reported above.

%-------------------------------------------------------------------------
%
% Section #2
%
%-------------------------------------------------------------------------
\section{The proposed line-clipping algorithm}
  \label{Sec: proposed}

\subsection{Methodology}
\begin{figure}[htb]
\centering
\includegraphics[width=.8\linewidth]{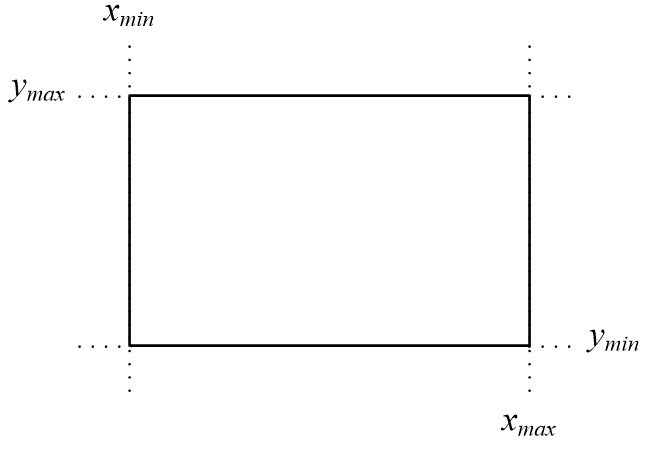}
\parbox[t]{.9\columnwidth}{\relax}
\caption{\label{fig:clippingRegion}Line-clipping region.}
\end{figure}

Assume that we want to clip a line inside a rectangle region or window that is defined by the points $(x_{\min}, y_{\max})$ and $(x_{\max}, y_{\min})$. 
This region is depicted in Fig.~\ref{fig:clippingRegion}.
Given two points $(x_1, y_1)$ and $(x_2, y_2)$ on the line that we want to clip, the slope $m$ of the line is constant and is defined by the ratio
\begin{equation}
  \label{Eq:Slope}
  m=\frac{y_2-y_1}{x_2-x_1}.
\end{equation}
For an arbitrary point $(x, y)$ on the line, the previous ratio can be written as
$$m=\frac{y-y_1}{x-x_1}.$$
Solving for $y$
$$y-y_1=m (x-x_1)\Leftrightarrow y = y_1 + m(x-x_1).$$
By replacing $m$ in this equation with Eq.~(\ref{Eq:Slope})
\begin{equation}
  \label{Eq:Solvey}
  y = y_1 + \frac{y_2-y_1}{x_2-x_1}(x-x_1).
\end{equation}
Solving for $x$, the equation becomes
\begin{equation}
  \label{Eq:Solvex}
  x = x_1 + \frac{x_2-x_1}{y_2-y_1}(y-y_1).
\end{equation}
Equations~(\ref{Eq:Solvey}) and (\ref{Eq:Solvex}) are two mathematical representations of the line equation $y=m x+b$ and will be used later by the algorithm in order to determine the part of the line that is inside the clipping window.

\subsection{The basic steps}
Suppose that the line which has to be clipped is defined by the points $(x_1, y_1)$ and $(x_2, y_2)$.
\subsubsection*{Step 1}
\begin{figure}[htb]
\centering
\includegraphics[width=.8\linewidth]{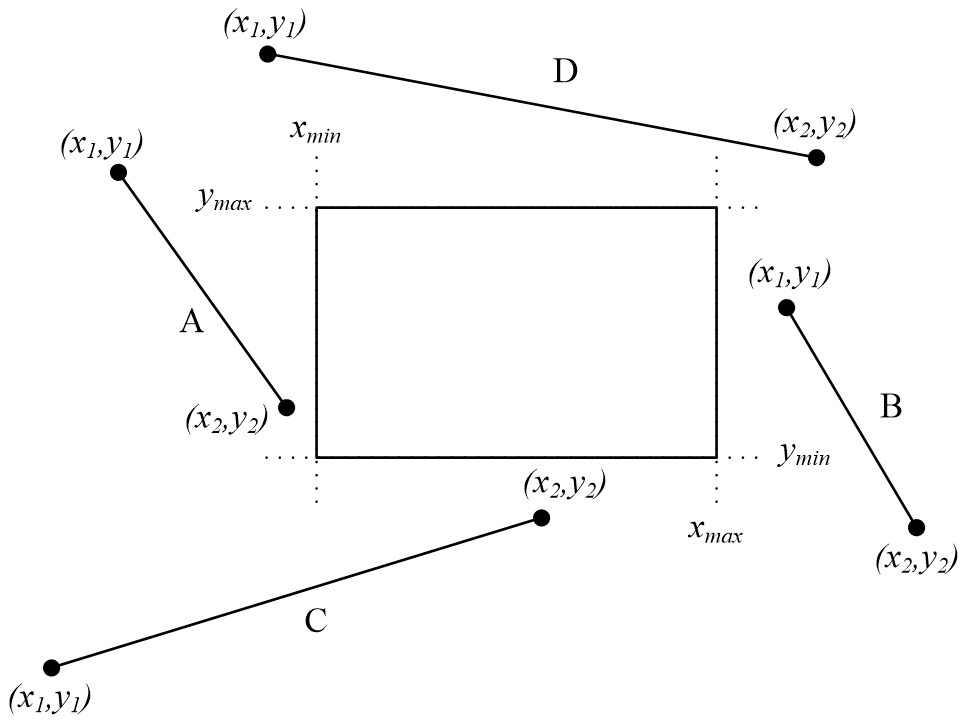}
\parbox[t]{.9\columnwidth}{\relax}
\caption{\label{fig:Rejection}Lines $A, B, C, D$ are rejected according to the first step of the algorithm.}
\end{figure}

The first step of the algorithm checks, if both points are outside the line clipping window and at the same time in the same region (top, bottom, right, left). 
If one of the following occurs, then the entire line is being rejected and the algorithm draws nothing (see Fig.~\ref{fig:Rejection}):

\begin{center}
\begin{tabular}{ l l }
$x_1<x_{\min}$ AND $x_2<x_{\min}$ & (line is left to the clipping window)\\
$x_1>x_{\max}$ AND $x_2>x_{\max}$ & (line is right to the clipping window)\\
$y_1<y_{\min}$ AND $y_2<y_{\min}$ & (line is under the clipping window)\\
$y_1>y_{\max}$ AND $y_2>y_{\max}$ & (line is over the clipping window)   
\end{tabular}
\end{center}
\subsubsection*{Step 2}
\begin{figure}[htb]
\centering
\includegraphics[width=.8\linewidth]{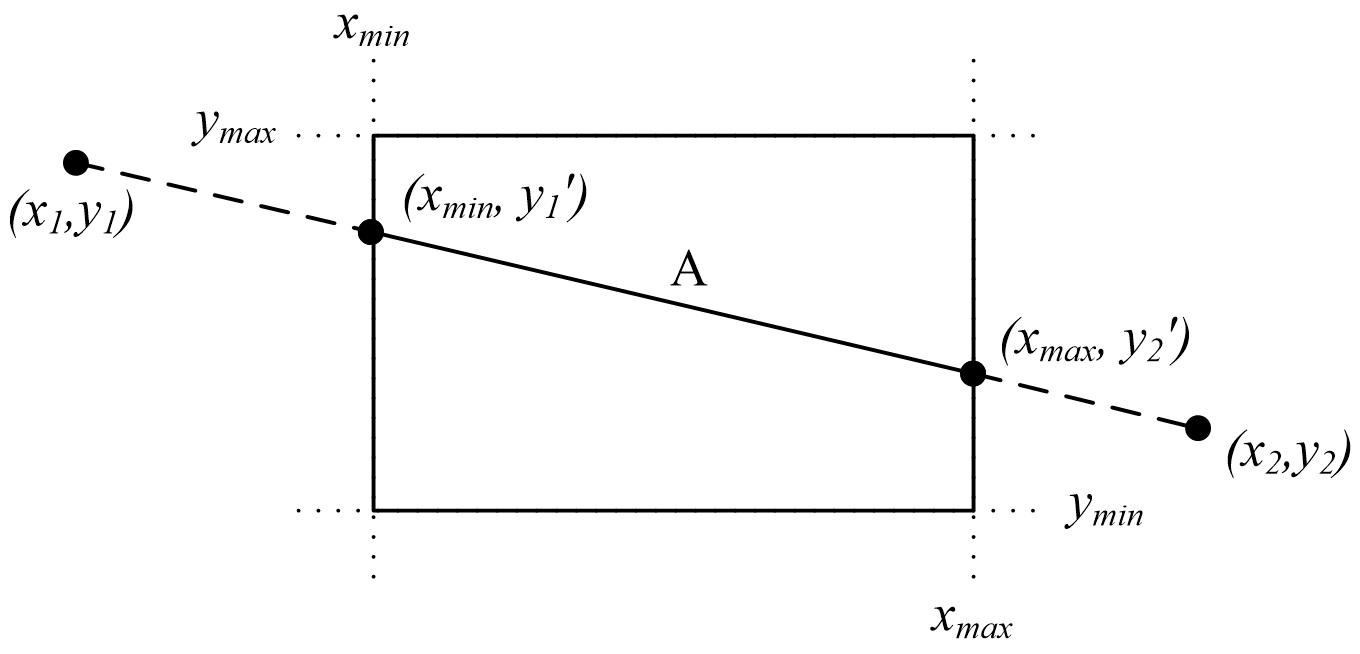}
\parbox[t]{.9\columnwidth}{\relax}
\caption{\label{fig:Clipping}Selecting the points of the line that are inside the clipping area.}
\end{figure}

In the second step, the algorithm compares the coordinates of the two points along with the boundaries of the clipping window. 
It compares each of the $x_1$ and $x_2$ coordinates with the $x_{\min}$ and $x_{\max}$ boundaries respectively, as well as each one of the $y_1$ and $y_2$ coordinates with the $y_{\min}$ and $y_{\max}$ boundaries. 
If any of these coordinates are out of bounds, then the specific boundary is used in the equation that determines the line in order to achieve clipping (see Fig.~\ref{fig:Clipping}).

For each of the coordinates of the two points and according to Equations~(\ref{Eq:Solvey}) and (\ref{Eq:Solvex}), the comparisons and changes made are:
\begin{itemize}
\item If $x_i<x_{\min}$, then
$$x_i=x_{\min}$$
$$y_i = y_1 + \frac{y_2-y_1}{x_2-x_1}(x_{\min}-x_1)$$
\item If $x_i>x_{\max}$, then
$$x_i=x_{\max}$$
$$y_i = y_1 + \frac{y_2-y_1}{x_2-x_1}(x_{\max}-x_1)$$
\item If $y_i<y_{\min}$, then
$$y_i=y_{\min}$$
$$x_i = x_1 + \frac{x_2-x_1}{y_2-y_1}(y_{\min}-x_1)$$
\item If $y_i>y_{\max}$, then
$$y_i=y_{\max}$$
$$x_i = x_1 + \frac{x_2-x_1}{y_2-y_1}(y_{\max}-x_1)$$
\end{itemize}
where $i=1, 2$.
\subsubsection*{Step 3}
The third and final step checks, if the new points, after the changes, are inside the clipping region and, if so, a line is being drawn between them.

\subsection{The algorithm in pseudo-code}

The representation of the algorithm in pseudo-code follows:

\begin{lstlisting}[basicstyle=\tiny][caption={Proposed algoritmh in pseudo-code}]

// x1, y1, x2, y2, xmin, ymax, xmax, ymin //

if not(x1<xmin and x2<xmin) and not(x1>xmax and x2>xmax) then
  if not(y1<ymin and y2<ymin) and not(y1>ymax and y2>ymax) then
    x[1]=x1
    y[1]=y1
    x[2]=x2
    y[2]=y2  
    i=1
    repeat
      if x[i] < xmin then
        x[i] = xmin
        y[i] = ((y2-y1)/(x2-x1))*(xmin-x1)+y1
      else if x[i] > xmax then
        x[i] = xmax
        y[i] = ((y2-y1)/(x2-x1))*(xmax-x1)+y1
      end if
      if y[i] < ymin then
        y[i] = ymin 
        x[i] = ((x2-x1)/(y2-y1))*(ymin-y1)+x1
        else if y[i] > ymax then 
        y[i] = ymax 
        x[i] = ((x2-x1)/(y2-y1))*(ymax-y1)+x1
      end if 
        i = i + 1 
    until i>2
    if not(x[1]<xmin and x[2]<xmin) and not(x[1]>xmax and x[2]>xmax) then 
        drawLine(x[1],y[1],x[2],y[2]) 
    end if
  end if
end if

\end{lstlisting}

%-------------------------------------------------------------------------
%
% Section #3
%
%-------------------------------------------------------------------------
\section{Evaluation of the proposed algorithm}
  \label{Sec:Eval}

\subsection*{Preparation}
In order to determine the efficiency of the proposed algorithm we decided to compare it with the six others: Cohen-Sutherland, Liang-Barsky, Cyrus-Beck, Nicholl-Lee-Nicholl, Skala and Kodituwakku-Wijeweere-Chamikara. 
%For being accurate and not to lead into misleading conclusions the comparison was made in two totally different programming languages: Scratch and C++ with OpenGL. In both environments, a basic aim for all of the algoritmhs was to have similar simple structure as well as they should be as efficient as possible. 

The combination of C++ and OpenGL was a good choice for evaluating the proposed algorithm for the following reasons: a) C++ along with OpenGL is a professional programming environment in computer graphics, b) OpenGL uses efficiently the computer's hardware as well as the graphics adapter, c) C++ is faster than many other programming languages, d) OpenGL is a portable language and the code can be tested easily in other operating systems or computers \cite{HBC14}.

\subsubsection*{The experiment}
\begin{figure}[htb]
\centering
\includegraphics[width=.8\linewidth]{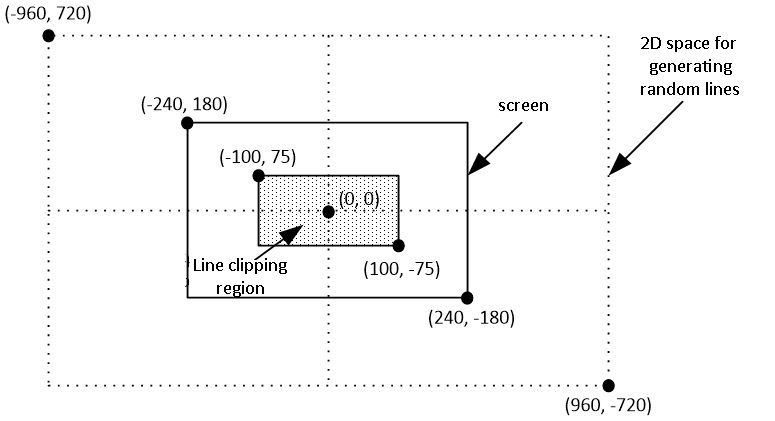}
\parbox[t]{.9\columnwidth}{\relax}
\caption{\label{fig:Space}Defining the 2D space for creating random line as well as definition of the line-clipping window.}
\end{figure}

The experiment was the following: Each one of the six evaluated algorithms would have to create a large number of arbitrary lines in a two-dimensional space. 
%The size of this 2D space should be four times larger than the Scratch screen for both environments. 
Such a space is determined by the points $(-960, 720)$ and $(960, -720)$. 
The line-clipping window should be at the centre of the screen and defined by the points $(-100, 75)$ and $(100, -75)$, in other words 200 pixels in width and 150 pixels in height (see Fig. \ref{fig:Space}). 
As someone may notice, the proportion of the screen and the clipping window is the same for both horizontal and vertical axis. 
The lines would be randomly generated anywhere in the 2D space and each algorithm would have to draw only the visible part of the lines inside the clipping window.

The time that each algorithm needs to clip and draw this large number of lines is recorded in every execution. 
The whole process is repeated 10 times and at the end the average time is being calculated.

\subsubsection*{Hardware and software specifications}
For realistic results, an average computer system was used for the experiment. The hardware as well as the software specifications were: a) Intel Core2Duo @ 2.60GHz CPU, b) RAM 2GB, c) AMD Radeon HD 5450 GPU, d) Windows 10 Professional operating system, e) Microsoft Visual C++ 2017 Enterprise Edition with OpenGL and glut v3.7 library.

\subsubsection*{Results}

We decided that each algorithm should draw 1,000,000 lines in every execution. 
The results are shown in Table~\ref{tab:executionInCpp}.

\begin{table}[h!]
\begin{center}
\caption{Execution times of each algorithm when creating 1,000,000 lines in C++ with OpenGL}
\label{tab:executionInCpp}
\begin{tabular}{|c|c|c|c|c|c|c|c|c|}
\hline
\textbf{Exec.} & \textbf{CS} & \textbf{LB} & \textbf{CB}& \textbf{NLN}& \textbf{Skala} & \textbf{KWC} & \textbf{Prop.}\\
\textbf{} & \textbf{(sec)} & \textbf{(sec)} & \textbf{(sec)} & \textbf{(sec)}& \textbf{(sec)} & \textbf{(sec)} & \textbf{(sec)}\\
\hline
\hline
1 & 1.302 & 1.264 & 1.446 & 1.577 &  1.234 & 1.216 & 1.182\\
\hline
2 & 1.362 & 1.233 & 1.425 & 1.376 & 1.313 & 1.224 & 1.125\\
\hline
3 & 1.453 & 1.225 & 1.471 & 1.437 &  1.299 &1.196 & 1.097\\
\hline
4 & 1.359 & 1.460 & 1.530 & 1.446 &  1.239 &1.271 & 1.176\\
\hline
5 & 1.407 & 1.263 & 1.519 & 1.455 &  1.337 &1.297 & 1.151\\
\hline
6 & 1.286 & 1.233 & 1.418 & 1.505 &  1.256 &1.268 & 1.216\\
\hline
7 & 1.295 & 1.182 & 1.439 & 1.427 &  1.352 &1.275 & 1.076\\
\hline
8 & 1.446 & 1.205 & 1.658 & 1.332 &  1.365 &1.223 & 1.209\\
\hline
9 & 1.420 & 1.218 & 1.423 & 1.448 &  1.248 &1.217 & 1.214\\
\hline
10 & 1.319 & 1.272 & 1.462 & 1.450 & 1.456 &1.251 & 1.199\\
\hline
\textbf{Avg:} & \textbf{1.365} & \textbf{1.256} & \textbf{1.479} & \textbf{1.445} & \textbf{1.310}& \textbf{1.244}& \textbf{1.165}\\
\hline
\end{tabular}
\end{center}
\end{table}

Finally, for a better evaluation of the previous results, the algorithms were executed again in C++ with OpenGL but this time 10,000,000 lines had to be clipped. The average time was noted down and the results can be seen in Table~\ref{tab:executionInCpp2}.

\begin{table}[h!]
\begin{center}
\caption{Execution times of each algorithm when creating 10,000,000 lines in C++ with OpenGL}
\label{tab:executionInCpp2}
\begin{tabular}{|c|c|c|c|c|c|c|c|c|}
\hline
\textbf{Exec.} & \textbf{CS} & \textbf{LB} & \textbf{CB}& \textbf{NLN} & \textbf{Skala}& \textbf{KWC} & \textbf{Prop.}\\
\textbf{} & \textbf{(sec)} & \textbf{(sec)} & \textbf{(sec)} & \textbf{(sec)} & \textbf{(sec)} & \textbf{(sec)}& \textbf{(sec)}\\
\hline
\hline
1 & 12.141 & 11.450 & 13.253 & 12.862 & 11.693 & 11.358 & 10.414\\
\hline
2 & 11.717 & 11.628 & 12.990 & 13.273 & 11.823 & 11.717 & 10.138\\
\hline
3 & 12.064 & 11.783 & 13.047 & 13.836 & 11.644 & 11.091 & 10.370\\
\hline
4 & 11.605 & 11.359 & 13.978 & 12.763 & 11.819 & 11.779 & 10.757\\
\hline
5 & 11.693 & 10.978 & 13.268 & 12.707 & 11.579 & 10.741 & 10.517\\
\hline
6 & 11.643 & 10.953 & 14.269 & 13.043 & 11.739 & 10.885 & 10.351\\
\hline
7 & 11.733 & 11.224 & 13.597 & 12.723 & 12.114 & 10.823 & 10.423\\
\hline
8 & 11.880 & 11.009 & 13.805 & 13.010 & 11.820 & 11.102 & 10.451\\
\hline
9 & 11.917 & 10.948 & 13.350 & 12.825 & 11.754 & 11.114 & 10.473\\
\hline
10 & 12.221 & 10.936 & 13.811 & 12.649 & 11.757 & 10.756 & 10.387\\
\hline
\textbf{Avg:} & \textbf{11.861} & \textbf{11.227} & \textbf{13.537} & \textbf{12.969} & \textbf{11.774} & \textbf{11.137}& \textbf{10.428}\\
\hline
\end{tabular}
\end{center}
\end{table}		

%-------------------------------------------------------------------------
%
% Section #4
%
%-------------------------------------------------------------------------
\section{Conclusions}
  \label{Sec:Conclusions}

\begin{figure}[htb]
\centering
\includegraphics[width=.9\linewidth]{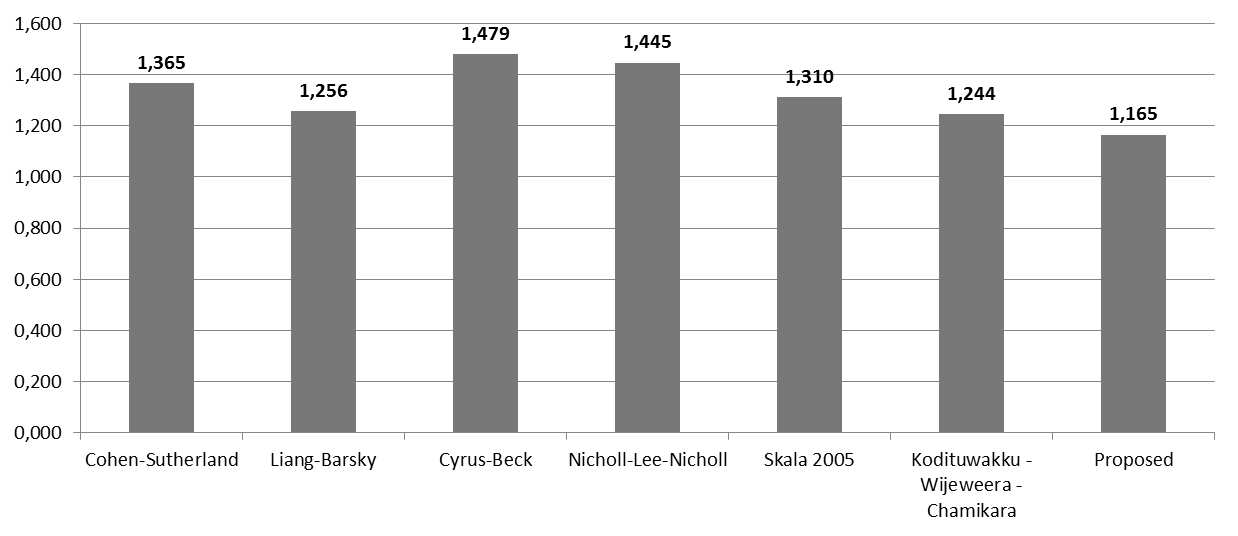}
\parbox[t]{.9\columnwidth}{\relax}
\caption{\label{fig:Comparison2}Graph with the average time of each algorithm for 1,000,000 in C++ with OpenGL (from lower to higher value).}
\end{figure}

\begin{figure}[htb]
\centering
\includegraphics[width=.9\linewidth]{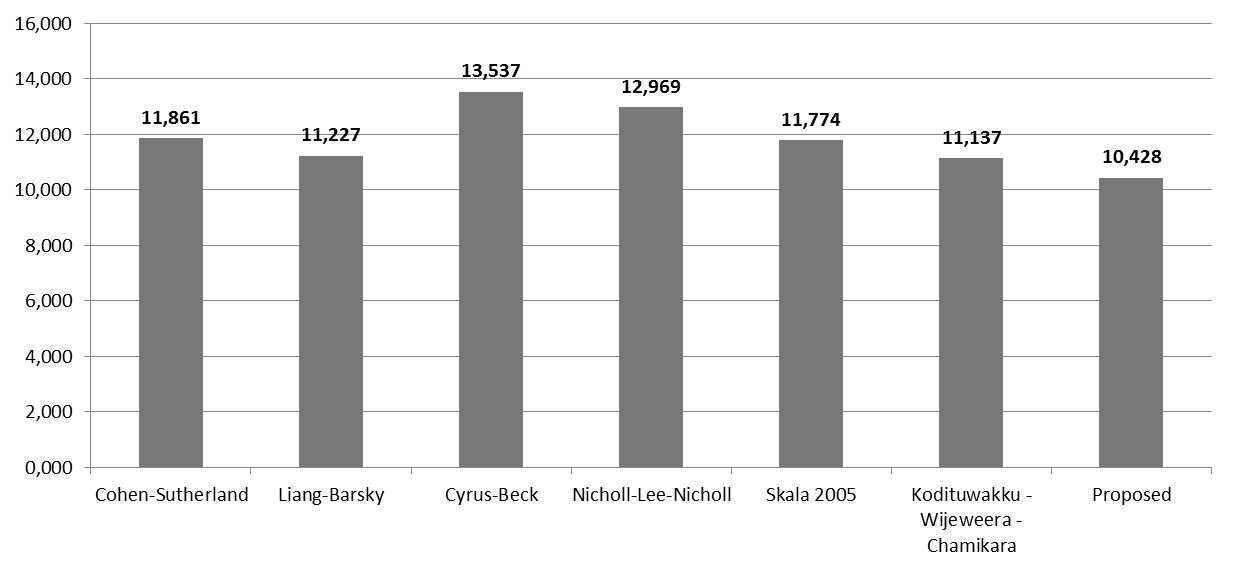}
\parbox[t]{.9\columnwidth}{\relax}
\caption{\label{fig:Comparison3}Graph with the average time of each algorithm for 10,000,000 in C++ with OpenGL (from lower to higher value).}
\end{figure}

In Fig.~\ref{fig:Comparison2} the graph of each case by using all data from the previous tables is illustrated. 
By using the average time of the algorithms executed in C++ with OpenGL for 1,000,000 lines and
\[
  \frac{|proposed - previous|}{proposed} \times 100
\]
we can see that that the proposed algorithm is 17.17\% faster than the Cohen-Sutherland, 7.81\% faster than the Liang-Barsky, 26.95\% faster than the Cyrus-Beck, 24.03\% than the Nicholl-Lee-Nicholl, 12.45\% than the Skala and 6.78\% faster than the Kodituwakku-Wijeweere-Chamikara algorithm. 
When the number of lines increases to 10,000,000, these time percentages are more or less the same; see Fig.~\ref{fig:Comparison3}. 
To be more specific, the proposed algorithm is 13.74\% faster than the Cohen-Sutherland, 7.66\% faster than the Liang-Barsky, 29.81\% faster than the Cyrus-Beck, 24.37\% than the Nicholl-Lee-Nicholl, 12.91\% than the Skala and 6.80\% faster than the Kodituwakku-Wijeweere-Chamikara algorithm.

As mentioned before, each algorithm has advantages and disadvantages. 
The Cohen-Sutherland algorithm is the oldest of all algorithms, it has an average performance comparing to the other five but it is difficult to implement in some programming environments, e.g. Scratch, due to the bitwise AND operations that it requires.

The Liang-Barsky algorithm performs very well and is almost as fast as the Kodituwakku-Wijeweere-Chamikara algorithm which was the faster algorithm after the proposed one. 
Liang-Barsky's main drawback is that it is slightly more difficult than the others to understand since it contains more advanced mathematical concepts.

The Cyrus-Beck algorithm looks like the Liang-Barksy, has the worst performance comparing to all the others and uses advanced mathematical concepts, too. 
But why should someone use this algorithm? 
A quick answer to this question could be: Because it can be modified very easily in order to clip polygons instead of lines.

The Nicholl-Lee-Nicholl algorithm is rather slow since it is faster only than the Cyrus-Beck. 
It uses a large number of subcases and subroutines for clipping a simple line and, as a result, it produces very long programs with many lines of code. 
An advantage of this algorithm is that it is easier to understand than others and thus to implement.

Finally, Skala as well as Kodituwakku-Wijeweere-Chamikara algorithm use a different approach than the popular ones, although fast they use a lot of conditions which make them more complicated and slower than the proposed one. 

%-------------------------------------------------------------------------
%
% Section #5
%
%-------------------------------------------------------------------------
\section{Summary}
\label{Sec:Sum}
There are many line-clipping algorithms in computer graphics. 
Each one has advantages and disadvantages. 
Cohen-Sutherland is the simplest line-clipping algorithm, but the Liang-Barsky algorithm is more efficient, since intersection calculations are reduced.
Overall, the afore-mentioned experimental results indicate that the proposed algorithm is simpler, faster and it certainly performs better than other known 2D line-clipping algorithms. 
It uses only a small number of variables and it is very easy to implement in any programming language or integrated development environment. 
Both the Cohen-Sutherland and Liang-Barsky algorithms can be extended to three-dimensional clipping. 
Nicholl-Lee-Nicholl cannot extend to three-dimensional clipping.
An interesting extension of the proposed algorithm would be clipping in three dimensions; see \cite{KWC12}, \cite{Nis17b}. 

%\section*{Compliance with Ethical Standards}

%Conflict of Interest: The authors declare that they have no conflict of interest.
%Funding: No funding was received.

\bibliography{IJCGA_final}
\bibliographystyle{plain}

%\vspace{2cm}

\section*{Authors}
\noindent {\bf Vasileios Drakopoulos} received a B.S. degree in Mathematics, an M.S. degree in Informatics \& Operations Research and a doctorate in Informatics and Computer Science from the National and Kapodistrian University of Athens, in 1990, 1992 and 1999, respectively. He began studying dynamic systems and fractals during his graduate studies and received a scholarship from the Bodossaki Foundation as financial support for doctorate studies. After completing his Ph.D., he received a Postdoctoral Scholarship from the (Greek) State Scholarships Foundation (I.K.Y) and has worked on parallel visualisation methods. He has taught a number of courses in Tertiary as well as in Secondary Education. Currently, he is an Assistant Professor in the Department of Computer Science and Biomedical Informatics at the University of Thessaly and a Research Fellow in the Department of Informatics \& Telecommunications at the National and Kapodistrian University of Athens. His scientific area of interests include Fractal and Computational Geometry, Computer Graphics, Dynamic Systems, Computational Complex Analysis, Image Processing and Compression, Human-Computer Interaction as well as Didactics of Informatics, Computer Science and ICT.\\

\noindent {\bf Dimitrios I. Matthes} graduated from the Department of Computer Systems, TEI of Piraeus in 2002 and obtained an M.S. diploma entitled ``Information Technology with Management'' in 2009. He has worked in many IT companies from 2002 until 2005. 
From 2005 onwards he works as a teacher of Informatics in secondary education. He has also worked as a part time teacher in several universities and institutes. 
Currently, he is pursuing his Ph.D. degree in Computer Science and Biomedical Informatics from the University of Thessaly. His research interests include Computer Graphics, Computational Geometry and ICT.

\end{document}